\documentclass[twocolumn,showpacs,preprintnumbers,amsmath,amssymb]{revtex4}

\usepackage{xcolor}
\usepackage{graphicx}
\usepackage{graphicx,epstopdf}
\usepackage{dcolumn}
\usepackage{bm}
\usepackage{mathrsfs} 

\begin{document}

\title{Resumption of dynamism in damaged networks of coupled oscillators}

\author{Srilena Kundu}
 \author{Soumen Majhi}
 \author{Dibakar Ghosh}
\affiliation{Physics and Applied Mathematics Unit, Indian Statistical Institute, 203 B. T. Road, Kolkata-700108, India}


\begin{abstract} 
Deterioration in the dynamical activities may come up naturally or due to environmental influences in a massive portion of biological and physical systems. Such dynamical degradation may have outright effect on the substantive network performance. This enforces to provide some proper prescriptions to overcome undesired circumstances. Through this article, we present a scheme based on external feedback that can efficiently revive dynamism in damaged networks of active and inactive oscillators and thus enhance the network survivability. Both numerical and analytical investigations are performed in order to verify our claim. We also provide a comparative study on the effectiveness of this mechanism for feedbacks to the inactive group or to the active group only. Most importantly, resurrection of dynamical activity is realized even in time-delayed damaged networks, which are considered to be less persistent against deterioration in form of inactivity in the oscillators. Furthermore, prominence in our approach is substantiated by providing evidence of enhanced network persistence in complex network topologies taking small-world and scale-free architectures, which makes the proposed remedy quite general. Besides the study in network of Stuart-Landau oscillators, affirmative influence of external feedback has been justified in network of chaotic R\"{o}ssler systems as well.

\end{abstract}


\maketitle


\section{Introduction}

Network theory offers an excellent platform to understand the universal properties of so many natural and engineered systems made up of a large number of units. Study of emergent collective behaviors in large ensembles of coupled dynamical units has got enormous appreciation because of its intensive applicability in imitating various self-organized complex systems \cite{sync1,sync2}. Among other perspectives, the one which deals with the exploration of network robustness (i.e., the ability to withstand perturbations) has extensive importance from several aspects. This scenario can be thought of in two different ways: {\it topological robustness} \cite{top1,top2,top3,top10,top5,top6,top7,top8,top9} and {\it dynamical robustness} \cite{dyn1,dyn2,dyn4,dyn5,dyn6,dyn7,dyn10,dyn13,dyn15,nw1,nw2}.The first one discusses about the persistence of network activities against structural perturbations in the form of removal of links (bond percolation) or nodes (site percolation) in the network. In this context, one of the most fundamental result demonstrates that the heterogeneous scale-free structures exhibit high resilience against random failure in the nodes whereas random networks are much less robust and the diameter of the network increases monotonically. But, scale-free networks are vulnerable to targeted attacks while for random connection topology, there is no significant change in the way of attacking owing to the homogeneity in the degree distribution \cite{top1,top3}. In contrast, for two interdependent networks, broader degree distribution (like that in scale-free structure) is much more sensitive to random failures \cite{top10,top6}.    
\par On the other hand, the second one of robustness is concerned with the network's survivability with respect to local perturbations in the dynamical activities of the nodes. This can be realized by exploring the evolution patterns of damaged networks made up of mixed populations comprising of active (healthy) and inactive (ill) dynamical units, known as {\it aging transition} \cite{dyn1} in the literature.  
 Infact, there exists several instances in ecological networks \cite{ecl1,ecl2,ecl3}, where some patches in the metapopulation become extinct that may have dramatic effects on the underlying developments. In neuronal networks, it is the rhythmicity of the neurons that governs the possibility of information exchange among them. So, loss in activity of a neuron may have several unexpected consequences \cite{new1}. Moreover, for appropriate functioning in cardiac and respiratory systems
 \cite{new2}, and specific physiological processes \cite{new3}, for instance cell necrosis within organs \cite{new4}, robust global oscillation is quite necessary.  Rather the study of dynamical robustness has also been extended to a network of complex topolgy \cite{nw1} or power grid network \cite{nw2} where the failure of a node is modeled via injecting noise into the dynamics of that node.
\par But despite of high relevance, the possible remedies to overcome the throughout dynamical failure of the network and hence to resurrect dynamism is yet to be fully explored and deserves significant consideration. The existing researches rather mainly focussed on the issue of aging transition under various interactional topologies of the network or using different coupling functions. For instance, such transitions are explained in globally \cite{dyn1,dyn2}, locally \cite{dyn4} coupled networks and in multilayer \cite{dyn5} networks as well. Crucial role of the low degree nodes in scale-free network \cite{dyn6} has also been discussed. Time-delay in the interactions may lower the network resilience under aging \cite{dyn10}.  Nevertheless, Liu et al. \cite{dyn13} put forward the notion of an additional parameter that controls the diffusion rate in order to enhance network persistence. Network robustness can also be developed by bringing uniform and normal random errors into the distance parameters of the system \cite{dyn15}.  In \cite{dyn7}, authors rendered a mechanism of recovering dynamical behavior in aging networks by additionally connecting supporting oscillators to the network. But adding intact oscillators in the network increases the effective size of the network. In the current article, we present an adaptable mechanism that involves introduction of external feedback in resurging dynamical activity in the network and hence developing network survivability, for which one does not need to change the intrinsic parameters of the system or to increase the effective network size.  
\par The concept of feedback is considered as one of the most important scientific understanding and as the heart of control theory \cite{fdb7,fdb5}. Particularly, positive feedback has been found to have colossal importance in natural systems \cite{fdb1} having impacts in evolutionary processes, physical systems, organism physiology, social evolutions, ecosystem and many more. It is also used in genetic networks \cite{fdb2}, neuronal networks \cite{fdbplos,fdb3} etc. In fact, positive feedback has been found to favor system instability in dynamical systems and utilized in elevating chaotic behavior and diverging from equilibrium, a scenario that we will be exploring in this work as well.
As far as the synchronization and control of networked dynamical systems are concerned, utility of feedback has been well justified \cite{fdb12,fdb13,fdb14,fdb15,fdb16,fdb17,fdb18,fdb21,fdb22,chaos2017p}. But, the influence of that entity in damaged networks of active (healthy) and inactive (diseased) dynamical systems is yet to be given attention, which is the focus of the present work. We put forward a detailed study on the ability of external positive feedback to improve the network survivability while the network is experiencing aging transition. We present analytical results on this issue that perfectly match the numerical ones. As discussed earlier \cite{dyn10}, time-delayed interaction among the systems may lower the network's persistence against local inactivation of the dynamical units. Here we show that even under this situation, feedback is the mechanism that is quite capable in enhancing network robustness. In addition, we explore this in complex topologies, such as small-world and scale-free networks that makes our idea independent of network architecture.   To demonstrate that our scheme is not system dependent, we provide results on both Stuart-Landau limit cycle system and chaotic R\"{o}ssler oscillator.
   
\par This paper is organized in the following way: In Section II, we provide a brief description of the network model of coupled Stuart-Landau oscillators. The general mathematical form of the network for globally interacting oscillators is provided in Section III. Numerical results followed by analytical study for globally interacting non-delay and delay coupled network are illustrated in Section IIIA and IIIB respectively. We show the effect of feedback in complex network topologies in Section IV. Section IVA deals with the results in case of small-world network, whereas, the results for scale-free network are summarized in Section IVB. Section V is devoted to the analysis of networked R\"{o}ssler systems. Finally, Section VI serves concluding remarks on the obtained results.

 \section{Model description of the damaged network of coupled Stuart-Landau Oscillators}

We consider the following network model of $N$ nodes as
\begin{equation}
	\begin{array}{lcl} \label{eq112}
\dot{z}_j = {\bf F}(z_j) + \frac{\epsilon}{N}\sum_{k=1}^N A_{jk}(z_k - z_j) + \eta f(\bar{z}),
\end{array}
\end{equation}
for $j= 1,2,...,N$ where ${\bf F}:\mathbb{R}^m\rightarrow\mathbb{R}^m$ represents the vector field corresponding to the system evolution whereas the function $f:\mathbb{R}^m\rightarrow\mathbb{R}^m$ defines the external feedback term. $A_{jk}$ is the adjacency matrix characterizing the connectivity pattern of the network, i.e.,  $A_{jk}=1$ if $j$-th and $k$-th nodes are connected and zero otherwise. The parameters $\epsilon$ and $\eta$ respectively accounts for the direct diffusive interaction strength and the strength of the feedback. 
\par Here we start by taking local dynamical units as the Stuart-Landau (SL) oscillators in the form \\
\begin{equation}
	\begin{array}{lcl} \label{eq113}
{\bf F}(z_j) =(\alpha_j + i\omega - |z_j|^2)z_j,
\end{array}
\end{equation}
for $j= 1,2,...,N$ where $\alpha_j$ are the internal parameters of the $j$-th system that define the distance from a Hopf bifurcation, $\omega$ is the natural frequency of oscillation and $i=\sqrt{-1}$. In isolation, $j$-th unit displays a stable limit cycle $\sqrt{\alpha_j} e^{i \omega t}$ if $\alpha_j>0$ and settles into the stable trivial fixed point $z_j=0$ for $\alpha_{j}\le 0$. 
As a result of this, active and inactive oscillators in the network respectively possess $\alpha_j=a>0$ and $\alpha_j=-b<0$. For the present work, we have chosen the external feedback to be linear of the form $f(\bar{z})= \bar{z}=\frac{1}{N}\sum_{k=1}^Nz_k$ (nonlinear feedback function could make the system more complicated without any additional advantage, infact this simple linear form is quite effective, as we have shown later). 
\par Next we follow the procedure as in \cite{dyn1} that defines the inactivation ratio (ratio of non self-oscillatory elements) $p$ which is the ratio of the number of inactive nodes and the total number of nodes in the network. Whenever $p$ exceeds a certain critical value $p_c$ (say), the global oscillation of the network dies out. So, our aim will be to restrain this sort of phase transition while utilizing the feedback parameter ($\eta$), as long as possible. We will explore the scenario of aging transition arising in the network in terms of the normalized order parameter $Z = \frac{|\bar{z}(p)|}{|\bar{z}(0)|}$, so that $|\bar{z}|$ identifies the intensity of global oscillation in the networked system and $Z$ is the normalized value of it.  We have fixed the network size $N=500$ (however, all the results are tested for larger networks). Without loss of generality, $a=2$, $b=1$ and $\omega=3$ are considered throughout the work \cite{mthd}.

\section{Globally coupled network}

In this section, we present a comprehensive study on the effect of feedback parameter to enhance dynamical robustness in the presence or absence of time-delayed interaction, while considering network of globally coupled oscillators. Then the mathematical form of the network reads as
\begin{equation}
	\small{\begin{array}{lcl} \label{eq116}
\dot{z}_j = (\alpha_j + i\omega - |z_j|^2)z_j + \frac{\epsilon}{N}\sum\limits_{k=1,k \neq j}^N [z_k(t-\tau) - z_j]+ \frac{\eta}{N}\sum\limits_{k=1}^Nz_k,
\end{array}}
\end{equation}   
where $\tau$ refers to the time delay in the direct interactions among the nodes. For the sake of simplicity, we set the group of active elements as $j \in \{1,2,...,N-Np\}$ and that of the  inactive elements as $j \in \{N-Np+1,...,N\}$.

\subsection{Non-delayed interaction}
Whenever there is no time delay i.e., with $\tau = 0$, the network \eqref{eq116} reduces to Eqn. \eqref{eq112} along with Eqn. \eqref{eq113}. But before going into the details of it, we first analyze the evolution in the dynamics of the oscillators before and after the inactivation procedure.
The dynamics of $N$ globally coupled synchronized Stuart-Landau oscillators for non-zero interaction strength, particularly for $\epsilon=3$ (with $\eta=0$), is depicted in Fig. 1, which is of limit cycle type with amplitude $\sqrt{2}$ for the active state (i.e., $p=0$), while in the inactive state (i.e., $p=1$) the nodes are stable at the trivial fixed point, i.e., the origin. The respective real parts of $z_j$, i.e., Re($z_j$) are shown in Figs. 1(a) and 1(b).  But considering a specific $p=0.2$ (i.e., when $20\%$ of the nodes are in inactive state), the inactive nodes start oscillating relying on the influence of the active nodes, which is depicted through Re($z_j$) in Figs. 1(c) and 1(d).

\begin{figure}[ht]
	\centerline{
		\includegraphics[scale=0.50]{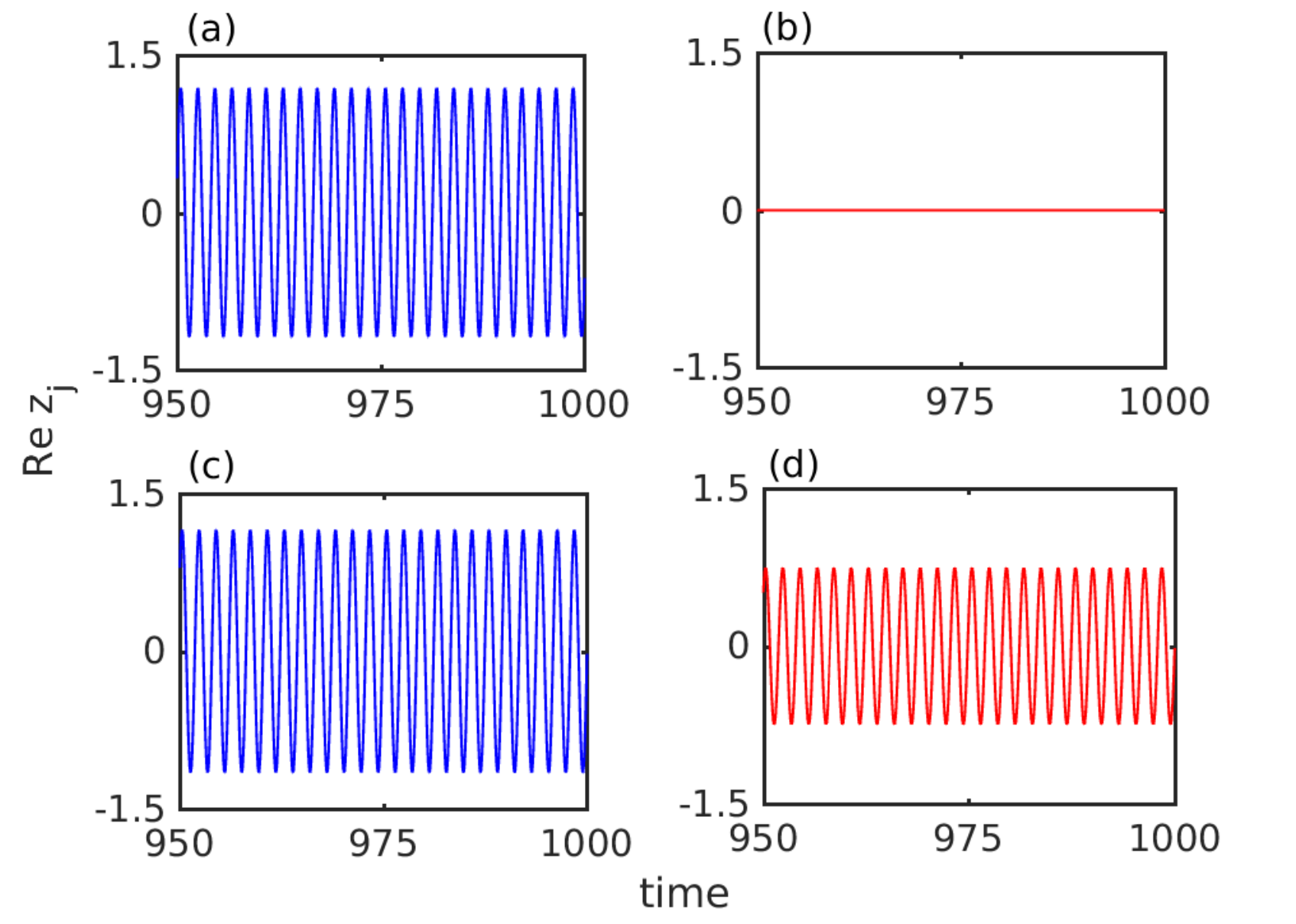}}
	\caption{The dynamics of $N$ completely synchronized globally coupled  Stuart-Landau oscillators in (a) active state, i.e., when $p=0$, and (b) inactive state, i.e., when $p=1$.  The behavior of (c) $(1-p)N$ active and (d) $pN$ inactive oscillators corresponding to the inactivation ratio $p=0.2$. Here, coupling strength $\epsilon=3$ and feedback parameter $\eta=0$. }
	\label{fig1}
\end{figure}

\par
\begin{figure}[ht]
	\centerline{
		\includegraphics[scale=0.5]{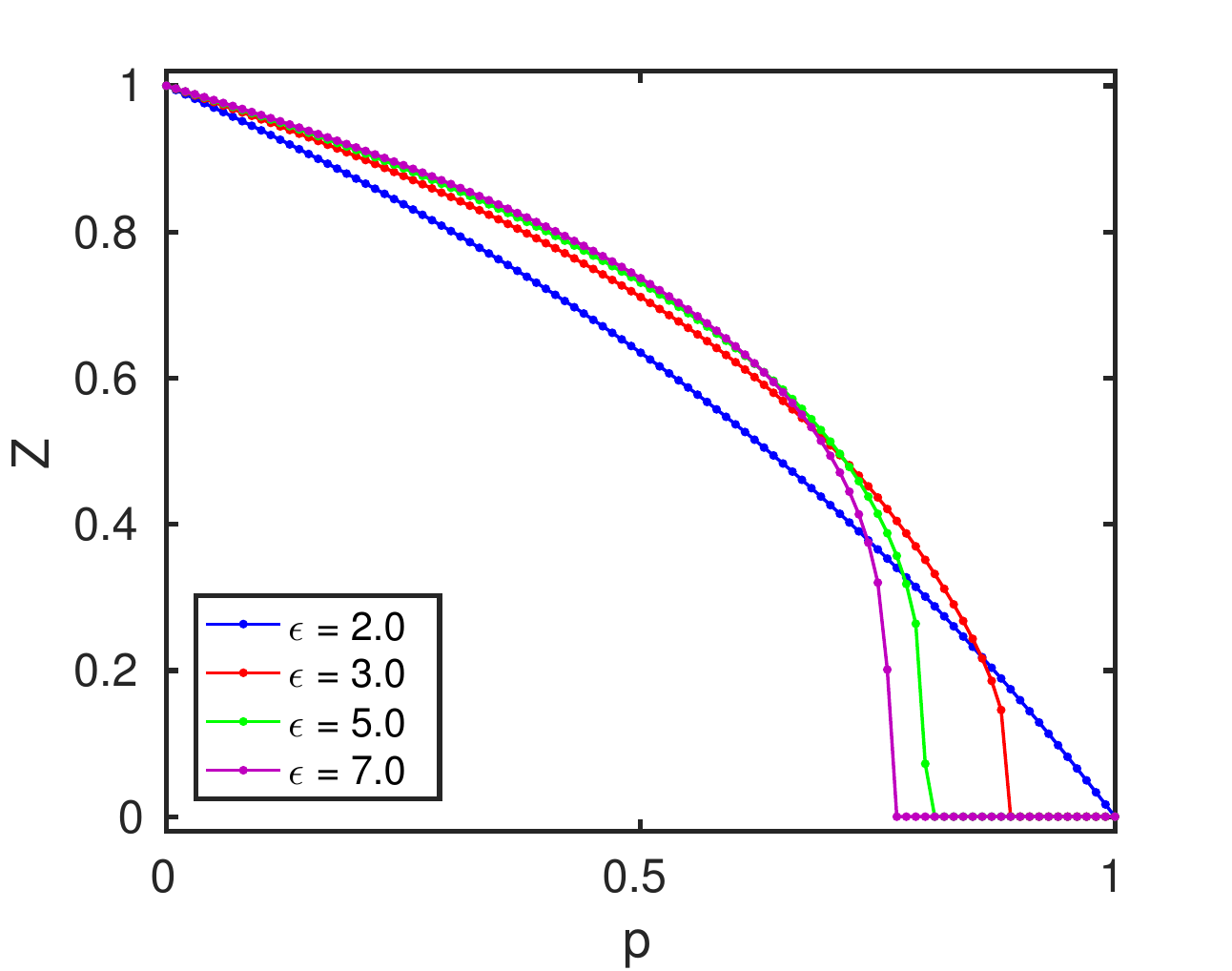}}
	\caption{The order parameter $Z$ vs. the inactivation ratio $p$ corresponding to various coupling strengths $\epsilon = 2, 3, 5$ and $7$ without feedback parameter ($\eta = 0$). The critical inactivation ratio $p_c$ decreases gradually as $\epsilon$ increases. }
	\label{fig2}
\end{figure}

 Figure \ref{fig2} shows the variation in $Z$ with respect to the increasing inactivation ratio $p$ $(0\le p\le 1)$ for different values of $\epsilon$ while $\eta=0$. As can be seen, aging transition (AT) occurs at $p_c=0.89$ (for $\epsilon=3.0$) where the order parameter $Z$ vanishes and the entire system gets stabilized to the trivial fixed point for $p\ge p_c$. However with smaller $\epsilon=2.0$, the transition appears only when all the nodes of network are made inactive, i.e., with $p_c=1$. Also for higher $\epsilon=5.0$  and $7.0$, the aging transition can be observed much earlier at $p_c=0.8$ and $p_c=0.77$ respectively. Infact, as the coupling strength $\epsilon$ increases, the critical inactivation ratio $p_c$ decreases \cite{dyn1}.  

\par
\begin{figure}[ht]
	\centerline{
		\includegraphics[scale=0.45]{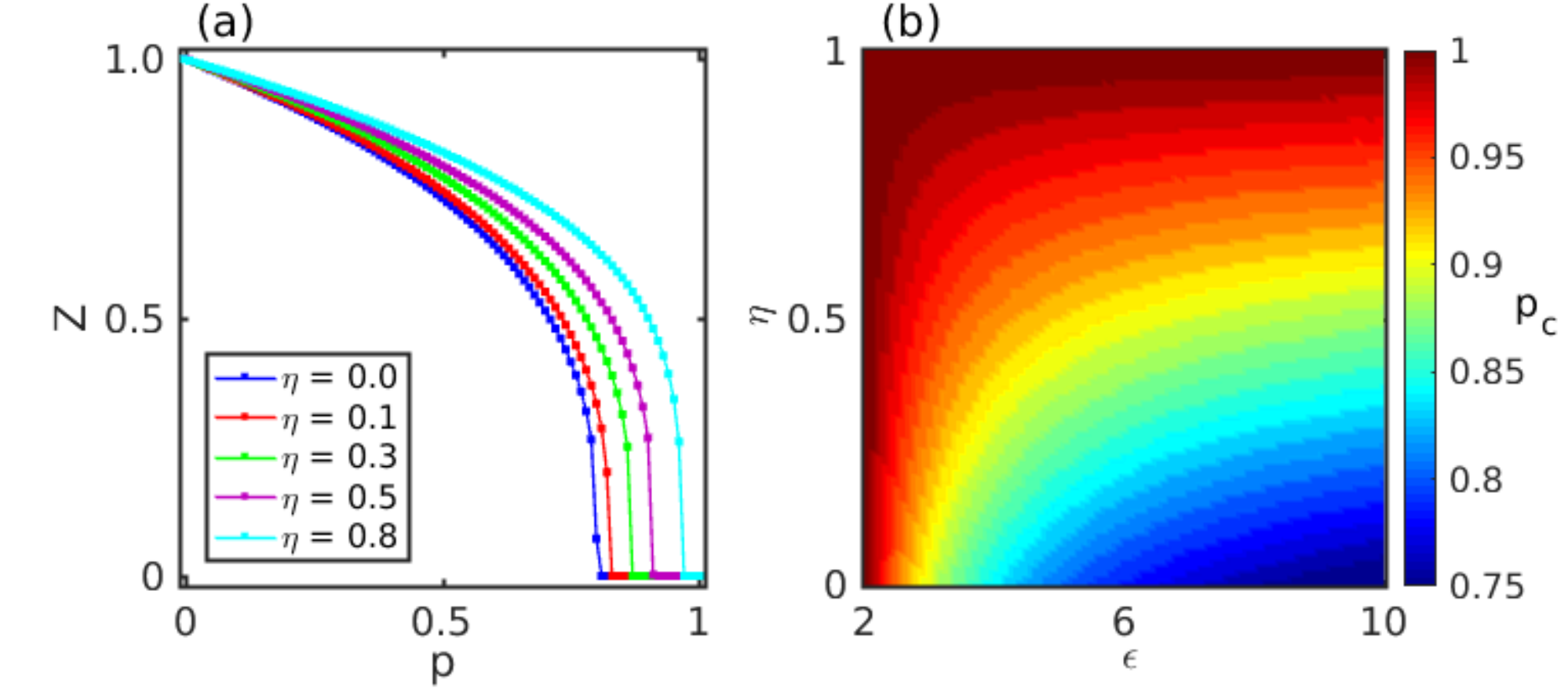}}
	\caption{(a) The order parameter $Z$ against the inactivation ratio $p$ corresponding to coupling strength $\epsilon = 5$ and various feedback parameter values $\eta = 0, 0.1, 0.3, 0.5$ and $0.8$. As $\eta$ increases $p_c$ value increases gradually. (b) Dependence of the critical ratio on coupling strength $\epsilon$ and feedback parameter $\eta$. The colorbar indicates the value of critical ratio $p_c$. }
	\label{fig3}
\end{figure}

 Now we will inspect the effects of introducing non-zero feedback strength $\eta$ into the network for fixed value of diffusive coupling strength $\epsilon$. For this, we first choose a definite value of $\epsilon=5.0$ and then see the diversity in the network dynamics by changing the feedback strength $\eta$. As shown earlier, $p_c$ is found to be $p_c\approx0.8$ when $\eta=0$. But a minute increment in $\eta$ to $\eta=0.1$ revives  the dynamical activity and hence enhance the network robustness to some extent, as $p_c$ increases to $p_c=0.82$ (cf. Fig. \ref{fig3}(a)). As we increase the value of $\eta$ to $\eta=0.3$ and $0.5$, the critical values $p_c$ become $p_c=0.87$ and $p_c=0.91$ respectively. This indicates a significant improvement in the resilience of the network to progressive dynamical inactivation of the nodes. Even higher $\eta=0.8$ leads the aging transition to occur at $p_c=0.97$. Thus for increasing $\eta$, the network is able to survive exhibiting global oscillation even when almost all the nodes are in inactive modes. For a better perception of this effect, next we plot the values of $p_c$ in the $\epsilon-\eta$ parameter plane, as in Fig. \ref{fig3}(b). The positive influence of increasing $\eta$ in resurgence of global oscillation for any $\epsilon$ (no matter how large) is conspicuous from the figure.   

\begin{figure}[ht]
	\centerline{
		\includegraphics[scale=0.55]{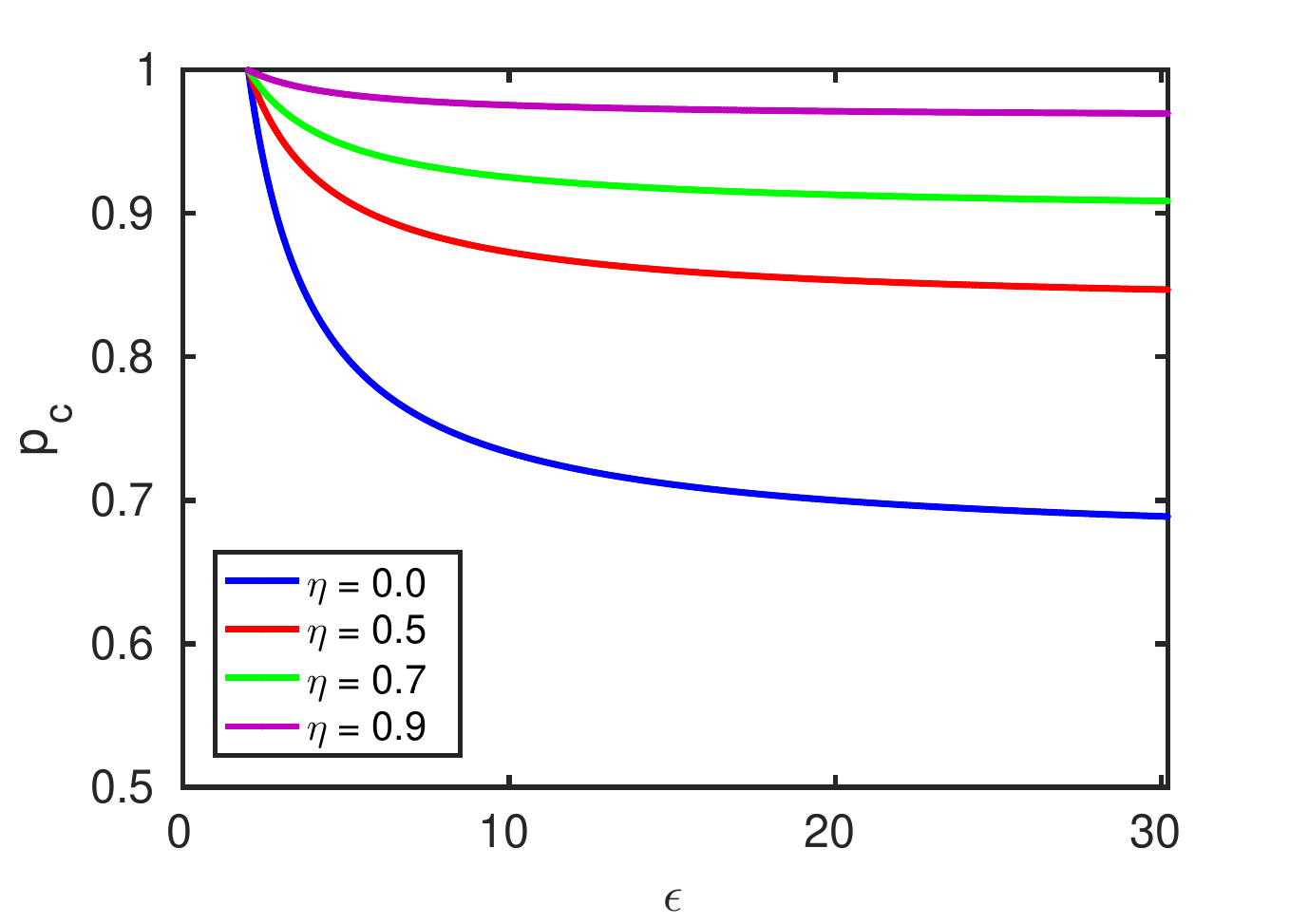}}
	\caption{The critical ratio $p_c$ vs. the coupling strength $\epsilon$ for different $\eta$ values obtained from the analytical Eqn. \eqref{eq115}. The dynamical robustness of the system \eqref{eq112} enhances with increasing $\eta$. }
	\label{fig4}
\end{figure}

Now to derive the critical value of inactivation ratio analytically, we assume $z_j = A (j =1,2,...,N-Np)$ for active oscillators and $z_j = I (j=N-Np+1,...,N)$ for inactive oscillators.  Then from Eqn. (\ref{eq116}) with $\tau = 0$, the reduced coupled system becomes 
\begin{equation}
	\begin{array}{lcl} \label{eq114}
	\dot{A} = (a + i\omega-\epsilon p+\eta q-|A|^2)A + (\epsilon+\eta) pI\\
	\dot{I} = (-b + i\omega-\epsilon q+\eta p-|I|^2)I + (\epsilon+\eta) qA,
	\end{array}
\end{equation}	
where $q=1-p$. Now as the aging transition correponds to the stabilization of the trivial fixed point $A=I=0$, we go through a linear stability analysis around origin that gives the following Jacobian matrix
$$\begin{pmatrix}
a +i\omega-\epsilon p+\eta q & (\epsilon+\eta)p \\
(\epsilon+\eta)q & -b +i\omega-\epsilon q+\eta p
\end{pmatrix}$$
The negativity of real parts of all the eigen values of this matrix characterizes the stability of the origin ($A=I=0$), investigation of which leads to the critical inactivation ratio $p_c$ as  
\begin{equation}
	\begin{array}{lcl} \label{eq115}
p_c = \frac{(a+\eta)(b+\epsilon)}{(\epsilon+\eta)(b+a)},

\end{array}
\end{equation}	
with $\epsilon \geq \epsilon_c = a$. Of course, for the case with no feedback, $p_c=\frac{a(b+\epsilon)}{\epsilon(b+a)}$ is the same as in \cite{dyn1}. This value of $p_c$ matches with the numerically calulated $p_c$ for $a,b,\epsilon$ and $\eta$ taken so far while genearting the Figs. \ref{fig2} and \ref{fig3}. The analytically obtained $p_c$ (cf. Eq. \ref{eq115}) with respect to increasing values of $\epsilon$ is figured out in Fig. \ref{fig4} for several values of $\eta$. The blue curve signifies the $p_c$ values against $\epsilon$ $(0\le \epsilon \le 30)$ whenever $\eta=0$. Initially, $p_c$ starts falling quite rapidly but after some $\epsilon \gtrsim 15$ , this fall is rather insignificant. As expected, a similar sort of drop in $p_c$ is observed whenever $\eta=0.5$ is chosen (the curve in red), but more importantly, this curve stays well above the previous one (in blue) indicating resumption of the dynamism in the network to a great extent irrespective of the interaction strength $\epsilon$. For higher $\eta=0.7$ and $0.9$, the curves (respectively in green and purple) depict even more enhancement in the network persistence. However, in all the cases after a certian $\epsilon$, minimal changes in the value of $p_c$ can be seen but AT is realized for any amount of coupling strength $\epsilon$. This scenario is in contrast to the $p_c-\epsilon$ variation reported in Ref. \cite{dyn13} where AT was observed only for finite intervals of coupling strength.

\begin{figure}[ht]
	\centerline{
		\includegraphics[scale=0.50]{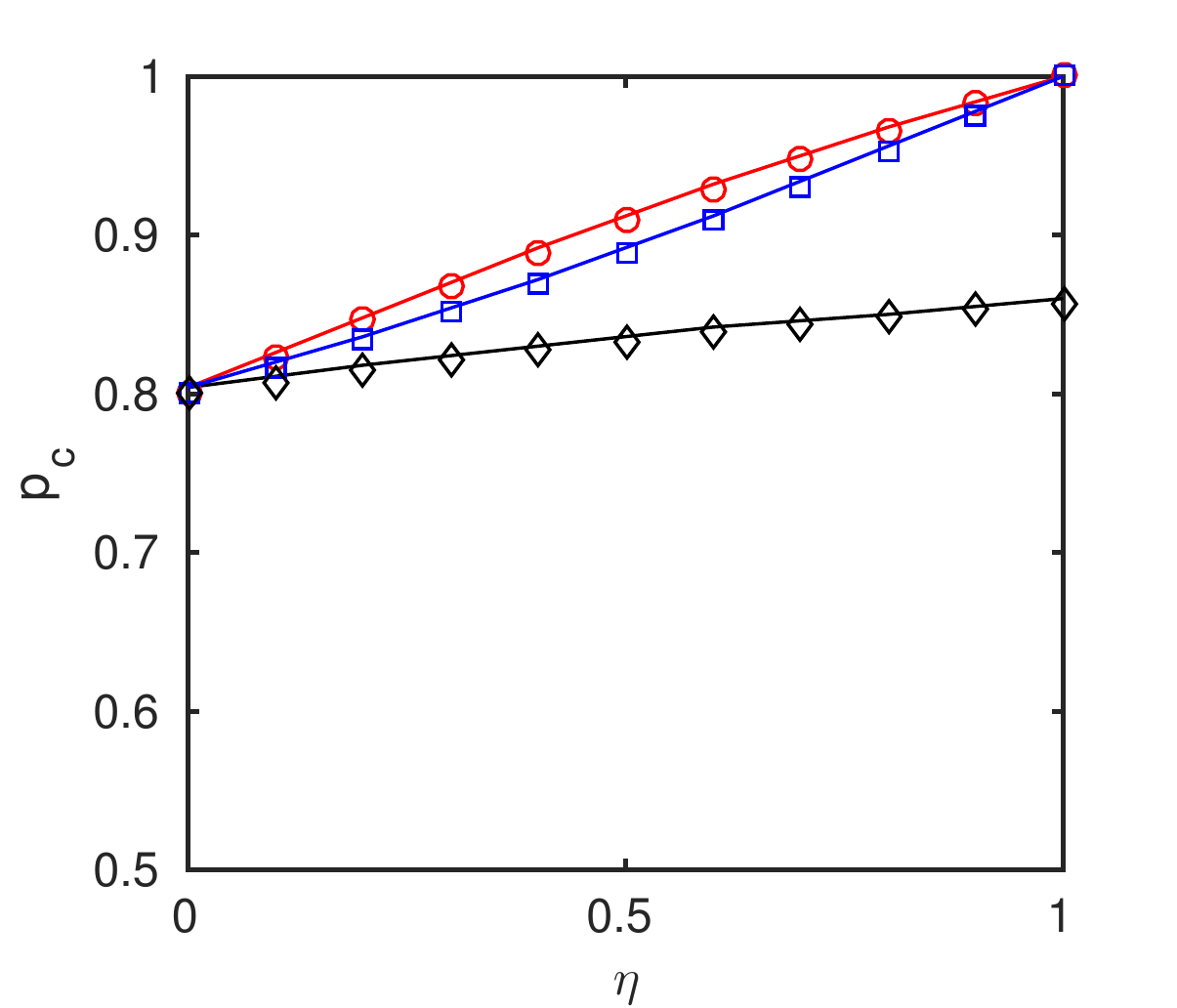}}
	\caption{The critical ratio $p_c$ against the feedback parameter $\eta$ for three different procedures of adding feedback to the network. Feedback is added to all the nodes(red), only to the inactive nodes(blue) and only to the active nodes(black). Here the coupling strength is fixed at $\epsilon = 5$. }
	\label{fig5}
\end{figure}

As we are dealing with a mechanism of inducing external feedback in a blended network ensemble of active and inactive dynamical systems, so one needs to scrutinize the impacts of different possible approaches of adding feedback to the network in detail. In order to do this, we plot the critical inactivation ratio $p_c$ versus the feedback strength $\eta$ while feedbacks are added to all the nodes, both active and inactive, in Fig. \ref{fig5} (in red, circles are analytical results whereas solid line corresponds to the numerical result). Since feedback is basically making the network more resilient (as discussed earlier), $p_c$ values monotonically increase with increasing $\eta$ ($0 \le \eta \le 1$). 
 After that, we employ feedback only to the nodes in inactive mode, for which the reduced system becomes 
\begin{equation}
	\begin{array}{lcl}
	\dot{A} = (a + i\omega-\epsilon p-|A|^2)A + \epsilon pI\\
	\dot{I} = (-b + i\omega-\epsilon q+\eta p-|I|^2)I + (\epsilon+\eta) qA,
	\end{array}
\end{equation}	
 and there exists meager difference in the $p_c$ values compared to the previous case. The analytical ($p_c=\frac{a(b+\epsilon)}{\epsilon(a+b)+\eta(a-\epsilon)}$) and numerical results \cite{op} are respectively shown by blue  squares and solid line. This implies that it may be enough to apply feedback only to the inactive nodes in order to enhance the dynamical survivability in the network. At the time of giving feedback to the active group of nodes only, the reduced system becomes 
\begin{equation}
	\begin{array}{lcl} 
	\dot{A} = (a + i\omega-\epsilon p+\eta q-|A|^2)A + (\epsilon+\eta) pI\\
	\dot{I} = (-b + i\omega-\epsilon q-|I|^2)I + \epsilon qA.
	\end{array}
\end{equation}	 
In which case $p_c=\frac{(a+\eta)(b+\epsilon)}{\epsilon(a+b)+\eta(b+\epsilon)}$ shows increasing feedback strength $\eta$ makes the network more robust but not as much as in the preceding two cases. The black diamonds and the solid line are below the red and blue ones, as in Fig. \ref{fig5}. This study helps one to perceive the way one should embed the feedback function in the system.  

\subsection{Time-delayed interaction and feedback}
\begin{figure}[ht]
	\centerline{
		\includegraphics[scale=0.45]{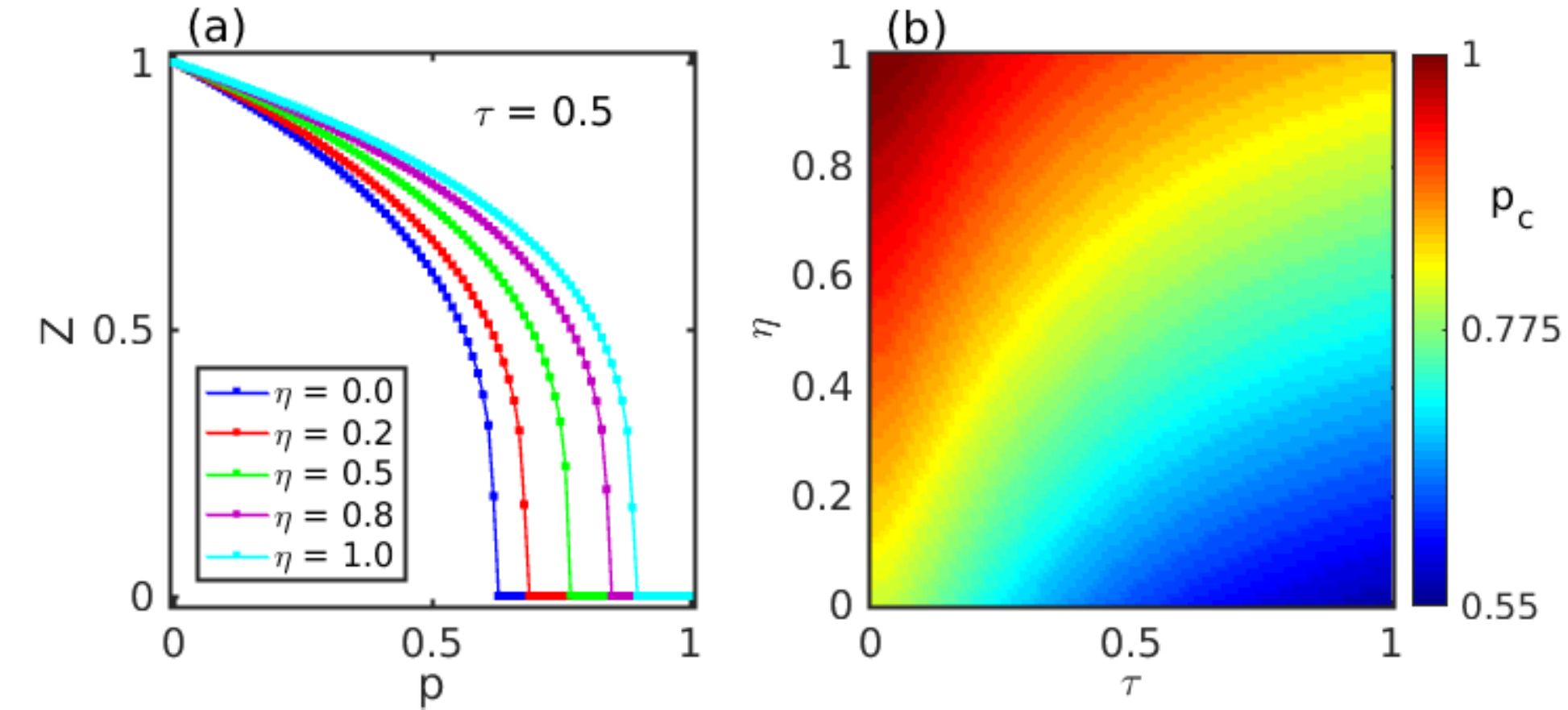}}
	\caption{(a) The order parameter $Z$ against the inactivation ratio $p$ for various $\eta$ values in the delay coupled network \eqref{eq116} with $N = 500$, $\tau = 0.5$ and $\epsilon = 5$. (b) Dependence of the value of critical ratio $p_c$ on $\eta-\tau$ parameter space. }
	\label{fig6}
\end{figure}

It is the purpose of this subsection to inspect what feedback does whenever there is time delay in the diffusive interaction among the globally coupled nodes.  According to the results addressed in \cite{dyn10}, $p_c$ decreases for increasing values of $\tau$ that readily suggests a de-enhancing tendency of delay, in absence of the feedback term. As reported in the previous subsection, for a fixed coupling strength $\epsilon=5$ and feedback strength $\eta=0$, without delay $\tau$, the value of $p_c$ happens to be $p_c=0.8$. But as delay $\tau=0.5$ is incorporated in the system, $p_c$ decreases to $p_c=0.63$, as displayed in Fig. \ref{fig6}(a). Remarkably, if $\eta$ is now introduced in the network, it becomes more resilient which can be easily discernible from the values of $p_c=0.69,0.77$ respectively for $\eta=0.2,0.5$. Higher $\eta=0.8,1.0$ are perfectly able to resume dynamical activity in the form of global oscillation to a greater extent as $p_c$ turns into $p_c=0.85,0.9$ respectively, as depicted in Fig. \ref{fig6}(a). Moreover, the impact of $\eta$ is illustrated for a sufficiently long range of the delay $\tau \in [0,~1]$ through the Fig. \ref{fig6}(b) that readily describes the enhancing effect of increasing $\eta$ for any value of $\tau$. Thus as observed in the instantaneous interaction scheme, here again $\eta$ is developing the network survivability altogether even in the presence of time delay in the coupling. 

\par On the other hand, following the similar approach as in previous case for globally coupled oscillators, the reduced model for active and inactive oscillators in case of time-delayed interaction with feedback mechanism becomes 
\begin{equation}\label{eq117}
\begin{array}{lcl}
\dot{A}(t) = (a+i\omega - \epsilon(1-\frac{1}{N}) - |A(t)|^2 +\eta q)A(t) + \eta p I(t)\\~~~~~~~~~~ +\epsilon(q-\frac{1}{N})A(t-\tau) + \epsilon p I(t- \tau),\\
\dot{I}(t) = (-b+i\omega - \epsilon(1-\frac{1}{N}) - |I(t)|^2 +\eta p)I(t) + \eta q A(t)\\~~~~~~~~~ +\epsilon(p-\frac{1}{N})I(t-\tau) + \epsilon q A(t- \tau).
\end{array}
\end{equation}
From linear stability analysis of Eq. \eqref{eq117} around the origin and setting the real part of the eigenvalue equal to zero, we obtain the characteristic equation for eigenvalues as 
\begin{equation}\label{eq118}
\begin{array}{lcl}
[a-\epsilon (1-\frac{1}{N})+i(\omega - \lambda_I)+\eta q +\epsilon (q-\frac{1}{N})e^{-i\lambda_I \tau}]\\ \times [-b-\epsilon (1-\frac{1}{N})+i(\omega - \lambda_I)+\eta p +\epsilon (p - \frac{1}{N})e^{-i\lambda_I \tau}]\\ - pq(\eta + \epsilon e^{-i\lambda_I \tau})^2 = 0,
\end{array}
\end{equation}           
where $\lambda_I$ is the imaginary part of the eigenvalue $\lambda$ i.e., $\lambda = i \lambda_I$. Separating real and imaginary parts we get the following equations
\begin{equation}\label{eq119}
\begin{array}{lcl}
[\omega - \lambda_I - B\sin(\lambda_I \tau)][\omega - \lambda_I - C\sin(\lambda_I \tau)]\\-[g_1+g_3+B\cos(\lambda_I \tau)][g_2+g_4+C\cos(\lambda_I \tau)]\\ = -g_3 g_4 - D_1 \cos(\lambda_I \tau) - D_2 \cos(2 \lambda_I \tau),
\end{array}
\end{equation}
and
\begin{equation}\label{eq120}
\begin{array}{lcl}
[\omega - \lambda_I - B\sin(\lambda_I \tau)][g_2+g_4+C\cos(\lambda_I \tau)]\\ + [\omega - \lambda_I - C\sin(\lambda_I \tau)][g_1+g_3+B\cos(\lambda_I \tau)]\\ = - D_1 \sin(\lambda_I \tau) - D_2 \sin(2 \lambda_I \tau),
\end{array}
\end{equation}
where $g_1=a-\epsilon(1-\frac{1}{N}), B = \epsilon(q-\frac{1}{N}), g_2 = -b-\epsilon(1-\frac{1}{N}), C = \epsilon(p-\frac{1}{N}), g_3 = \eta q, g_4 = \eta p, D_1 = 2 \eta \epsilon pq$ and $D_2 = \epsilon^2 pq$.

\begin{figure}[ht]
	\centerline{
		\includegraphics[scale=0.45]{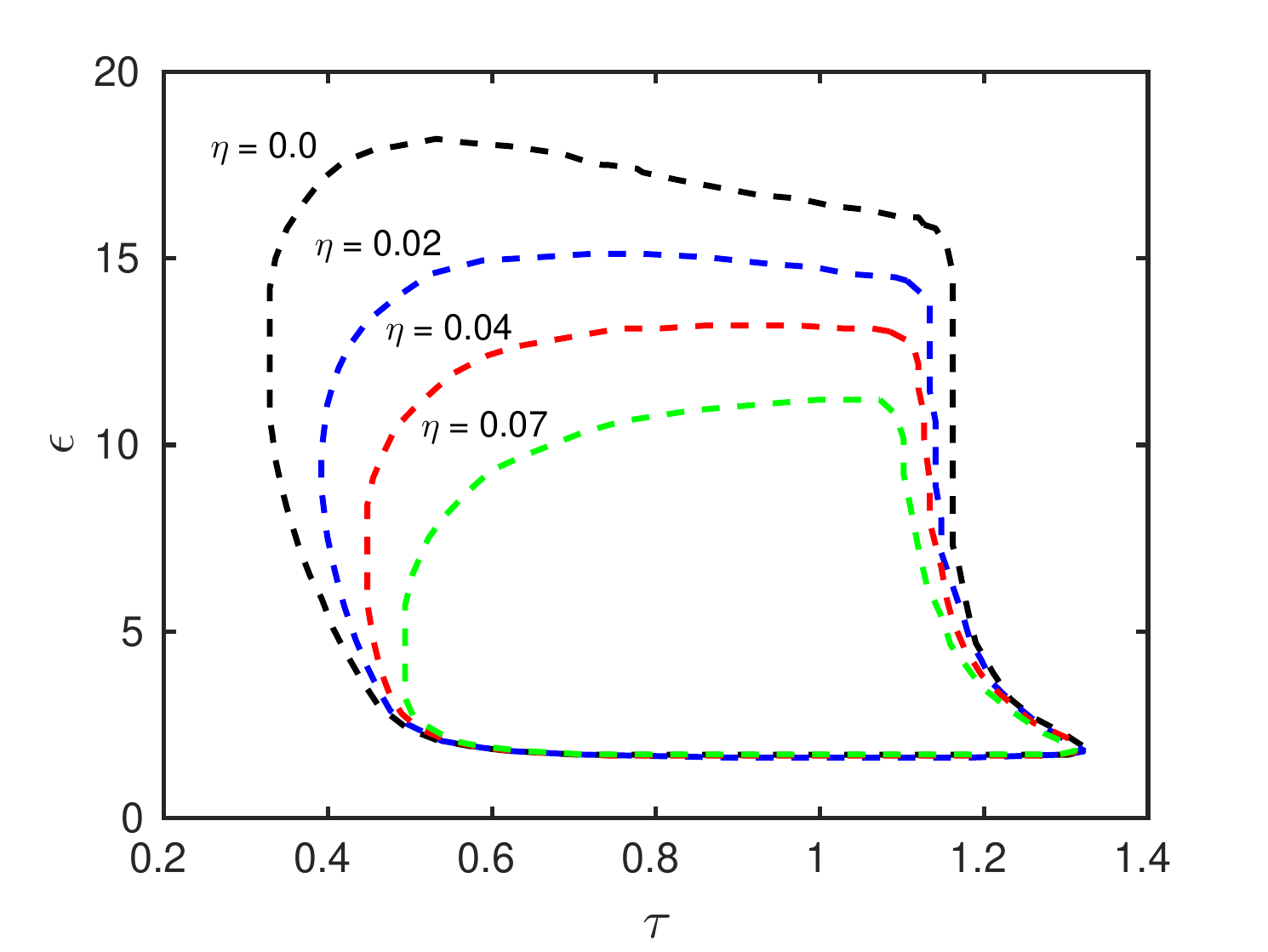}}
	\caption{Aging islands in the $\epsilon - \tau$ parameter plane at different values of the feedback parameter $\eta$ for fixed inactivation ratio $p = 0.65$. Here the size of the aging islands reduces significantly with the increase of $\eta$ value.}
	\label{fig7}
\end{figure}

\par Figure \ref{fig7} depicts the aging islands in the $\epsilon-\tau$ parameter plane obtained from the set of equations (\ref{eq119}) and (\ref{eq120}) for different values of the feedback strength $\eta$. For a fixed inactivation ratio $p=0.65$, firstly the aging island is plotted (in black) whenever $\eta=0$. Next the same is displayed for $\eta=0.02$ (in blue) in Fig. \ref{fig7}. It is easily observed that the aging island in the parameter plane gets reduced significantly for this non-zero $\eta$ that readily implies enhancement in the network's dynamical persistence. Further increment in $\eta$ to $\eta=0.04$ and $0.07$ (in red and green respectively) helps in shortening the aging island area more comprehensively.

\section{Interaction on complex networks}

This section is devoted to the discussion of efficiency of the external feedback function on developing dynamical robustness in a network possessing complex interactional topologies. Particularly, we will be analyzing this scenario on top of both small-world and scale-free architectures. 

\subsection{Small-world interaction}

Small-world networks \cite{swn} appear as a result of random rewiring of links (with a certain probability $p_{sw}$) in a regular (lattice) network, interpolating between two limiting cases of regular ($p_{sw}=0$) and random ($p_{sw}=1$) topologies that maintains low diameter and high clustering coefficient in the network. 
\par
\begin{figure}[ht]
 	\centerline{
 		\includegraphics[scale=0.48]{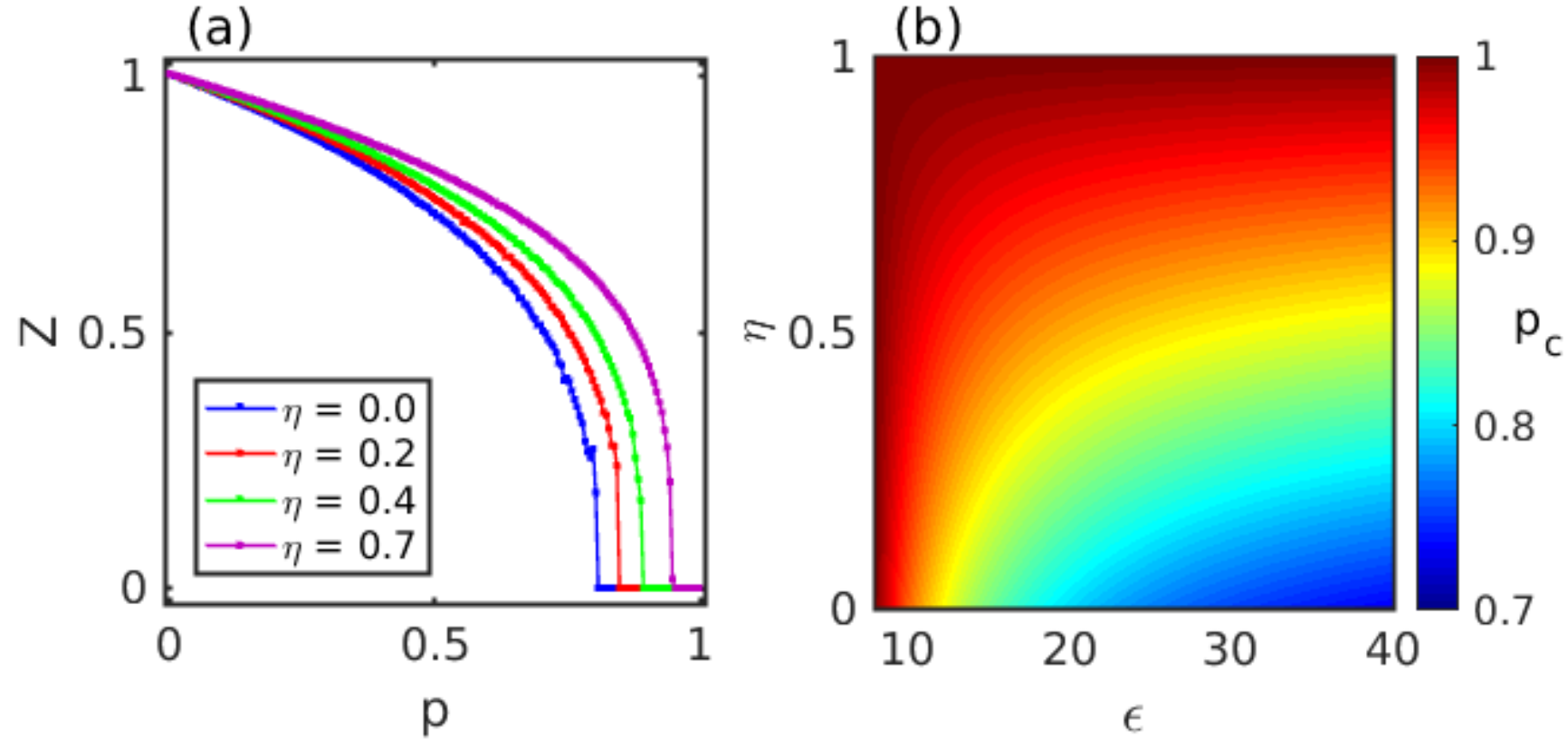}}
 	\caption{Effect of feedback parameter on small-world network: (a) The order parameter $Z$ against the inactivation ratio $p$ for various $\eta$ values in small-world network with  $d=0.24$ and coupling strength $\epsilon=20$, (b) Dependence of the critical ratio $p_c$ in $\epsilon-\eta$ parameter space.}
 	\label{fig8}
 \end{figure}

 Choosing $N=500$, $p_{sw}=0.01$ and the link density $d=\langle k \rangle/(N-1)=0.24$ ($\langle k \rangle$ being the average degree of the nodes). Figure \ref{fig8}(a) shows variation in the order parameter $Z$ against $p$ for $\eta=0,~0.2,~0.4$ and $0.7$ while the oscillators are interacting over a small-world topology. In absence of feedback, $Z$ declines to zero at $p_c=0.805$ indicating aging transition for coupling strength $\epsilon=20$. But, $p_c$ value increases to $p_c=0.85$ and $p_c=0.892$ respectively with $\eta=0.2$ and $\eta=0.4$ signifying improvement in the network survivability. Whenever $\eta=0.7$, the network becomes more resilient as it turns out to be able to retain dynamism even upto $p=0.95$.
\par Next we analytically derive the critical ratio $p_c$ for random inactivation in small-world network in presence of the feedback parameter $\eta$. 
The reduced model for active and inactive oscillators in case of small-world network with feedback parameter $\eta$ can be written as 
\begin{equation}
	\begin{array}{lcl} \label{eq121}
	\dot{A} = (a + i\omega-\epsilon d p+\eta q-|A|^2)A + (\epsilon d+\eta) pI\\
	\dot{I} = (-b + i\omega-\epsilon d q+\eta p-|I|^2)I + (\epsilon d+\eta) qA,
	\end{array}
\end{equation}
as the number of inactive oscillators in the neighborhood of each oscillator is expected to be $p\langle k \rangle$ and that of the active oscillators is $(1-p)\langle k \rangle$, where $q=1-p$ and $d = \langle k \rangle/N-1$ is the link density of the network. From the linear stability analysis around the fixed point $A=I=0$, we get the critical inactivation ratio $p_c$ as 
\begin{equation}
	\begin{array}{lcl} \label{eq122}
p_c = \frac{(a+\eta)(b+\epsilon d)}{(\epsilon d+\eta)(b+a)},

\end{array}
\end{equation}	
with $\epsilon \geq \epsilon_c = a/d$.

\par Having this value of the critical inactivation ratio $p_c$ (cf. Eqn. \ref{eq122}), we plot its dependence on simultaneous variation of the coupling strength $\epsilon \in [8,~40]$ and feedback parameter $\eta \in [0,~1]$ for small-world network in Fig. \ref{fig8}(b). From the figure, it is quite conspicuous that higher the feedback strength more the network is dynamically persistent, irrespective of the strength of interaction $\epsilon$.

\subsection{Scale-free interaction}

 On the other hand, a scale-free architecture \cite{sfn} corresponds to a highly heterogeneous sceanrio as far as the degree distribution is concerned that basically follows a power-law $P(k)\sim k^{-\gamma}$, where $P(k)$ is
the probability of finding a node of degree $k$ and $\gamma$
 is the power-law exponent (in our case, $\gamma=3.0$).
 We next study this transition for a scale-free network of $N=500$ oscillators and $d=0.078$. As can be observed from Fig. \ref{fig10}(a), with a fixed interaction strength $\epsilon=200$ and no feedback, the order parameter drops down to zero for $p\ge p_c=0.723$. Importantly enough, with a non-zero feedback $\eta=0.2$, the value of $p_c$ increases to $p_c=0.782$. In a similar fashion, higher feedback strengths $\eta=0.5$ and $\eta=0.8$ lead to highly improved crtitical inactivation ratios $p_c=0.867$ and $p_c=0.95$ respectively. These outcomes are the indicators of resumption of dynamism in damaged complex networks of active and inactive units.

Again, according to the degree-weighted mean field approximation \cite{dwmf}, the original local field can be approximated as 
\begin{equation}
\begin{array}{lcl}
f_{org_j} = \sum\limits_{k=1}^N A_{jk}z_k \simeq (1-p)k_jH_A(t) + pk_jH_I(t) = f_{app_j},
\end{array}
\end{equation}
where $H_A(t) = \frac{\sum_{j\in S_A} k_jz_j(t)}{\sum_{j\in S_A} k_j}$ and $H_I(t) = \frac{\sum_{j\in S_I} k_jz_j(t)}{\sum_{j\in S_I} k_j}$ are the degree-weighted mean fields for active and inactive groups of dynamical units respectively and $k_j(j=1,2,...,N)$ is the degree of the $j$-th node. Though the scale-free network is highly heterogeneous as far as the degree distribution is
concerned, since we are actually dealing with degree-weighted mean fields $H_A(t) , H_I(t)$ that involves a
normalization through the degrees, so, this mean-field approach $f_{app_j}$ yields quite a good approximation for the local fields $f_{org_j}$ arising from a scale-free dynamical network as well. We justify this claim in Fig. 9 while plotting the approximated results of several oscillators  through mean-field approach that matches the original local fields.
\begin{figure}[ht]
 	\centerline{
 		\includegraphics[scale=0.5]{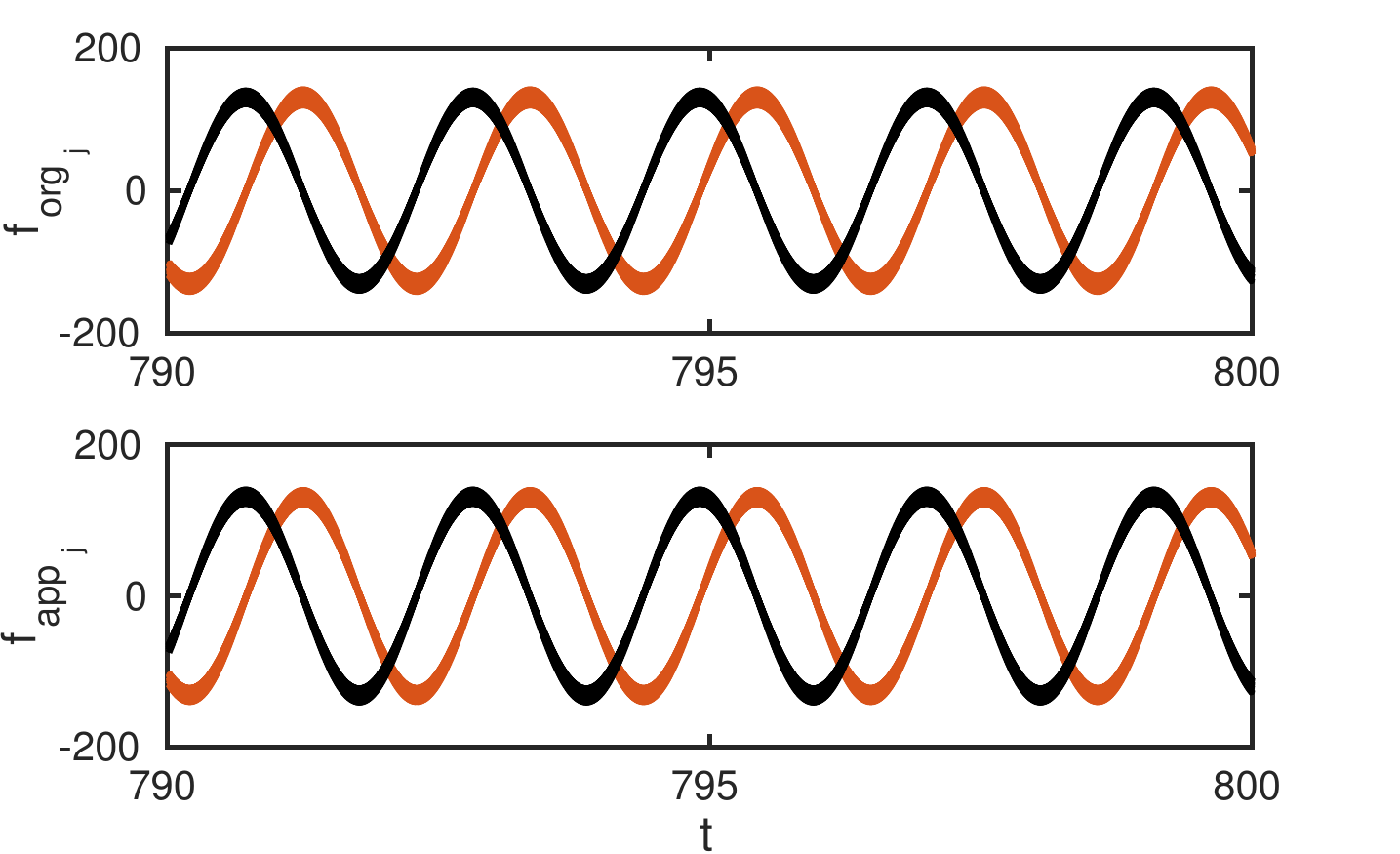}}
 	\caption{Validation of the original fields $f_{org_j}$ by the approximated mean-fields $f_{app_j}$ considering $N = 500, d = 0.078, \epsilon = 80$ and $p=0.5$. The upper panel shows the original local fields of several oscillators while the lower panel shows the corresponding approximated fields. Here black and brown color respectively corresponds to the real and imaginary parts of the local fields.}
 	\label{fig9}
 \end{figure}

Thus the original system \eqref{eq112} can be approximated as
\begin{equation}
\begin{array}{lcl} \label{eq123}
\dot{z_j} = (\alpha + i\omega - |z_j|^2)z_j + \frac{\epsilon k_j}{N}((1-p)H_A(t)+pH_I(t)-z_j)\\~~~~~~ + \eta((1-p)H_A(t)+pH_I(t)).
\end{array}
\end{equation}
 Assuming that the state variables can be written as $z_j(t)=r_j(t)e^{i(\omega t+\theta)}$, where $r_j$ is the amplitude and $\theta$ is the phase shift and substituting it in Eq. \eqref{eq123} we obtain,
\begin{equation}
\small{\begin{array}{lcl}  \label{eq124}
\dot{r_j} = (\alpha - \frac{\epsilon k_j}{N}-r_j^2)r_j + (\frac{\epsilon k_j}{N}+\eta)((1-p)R_A(t)+pR_I(t)),
\end{array}}
\end{equation}
\\
where $R_A(t) = \frac{\sum_{j\in S_A} k_jr_j(t)}{\sum_{j\in S_A} k_j}$ and $R_I(t) = \frac{\sum_{j\in S_I} k_jr_j(t)}{\sum_{j\in S_I} k_j}$. Assuming the time-independence of $R_A(t)$ and $R_I(t)$ in the stationary oscillatory regime, the phase transition from oscillatory ($R_A,R_I>0$) to non-oscillatory ($R_A=R_I=0$) takes place due to the change in stability of the fixed point at the origin. The stability is determined by the following jacobian matrix 
$$J_0=\begin{pmatrix}
\frac{\partial G_A(R_A,R_I)}{\partial R_A} & \frac{\partial G_A(R_A,R_I)}{\partial R_I} \\
\frac{\partial G_I(R_A,R_I)}{\partial R_A} & \frac{\partial G_I(R_A,R_I)}{\partial R_I}
\end{pmatrix}\Biggr\rvert_{R_A=R_I=0},$$
where
\begin{equation}
\begin{array}{lcl}  \label{eq125}
G_A(R_A,R_I)=\frac{\sum_{j\in S_A}k_j r_j^*(R_A,R_I)}{\sum_{j\in S_A}k_j},\\\\
G_I(R_A,R_I)=\frac{\sum_{j\in S_I}k_j r_j^*(R_A,R_I)}{\sum_{j\in S_I}k_j}
\end{array}
\end{equation} 
and the stationary amplitude $r_j^*$ is given by a positive real solution of the following equation:
\begin{equation}
\begin{array}{lcl} \label{eq126}
r_j^3 - (\alpha_j-\frac{\epsilon k_j}{N})r_j - (\frac{\epsilon k_j}{N}+\eta)((1-p)R_A+pR_I)=0.
\end{array}
\end{equation}
Equation \eqref{eq126} has only one positive real root if we assume $\alpha_j - \frac{\epsilon k_j}{N} <0$ for all $j \in S_A$. Differentiating Eqns. \eqref{eq125} and \eqref{eq126} with respect to $R_A$ we obtain the first entry of $J_0$ as follows:
\begin{equation}
\begin{array}{lcl} \label{eq127}
\frac{\partial G_A}{\partial R_A}\Bigr\rvert_{R_A=R_I=0} = \frac{(1-p)\epsilon}{\sum_{j \in S_A}k_j} \left(\frac{1}{N}\sum_{j \in S_A}\frac{k_j^2}{\epsilon k_j/N - \alpha_j}\right) \\~~~~~~~~~~~~~~~~~~~~~ + \frac{(1-p)\eta}{\sum_{j \in S_A}k_j} \left(\frac{1}{N} \sum_{j \in S_A} \frac{k_j}{\epsilon k_j/N - \alpha_j}\right)\\\\~~~~~~~~~~~~~~~~~~ \simeq \frac{1}{d}\left(\frac{1}{N}\sum_{j\in S_A}\frac{d_j^2}{d_j-\alpha_j/\epsilon}\right) \\~~~~~~~~~~~~~~~~~~~~~ + \frac{\eta}{d \epsilon}\left(\frac{1}{N}\sum_{j \in S_A}\frac{d_j}{d_j-\alpha_j/\epsilon}\right), 
\end{array}
\end{equation}
where $d_j=k_j/N$ is the ratio of the degree of $j$-th oscillator and the system size, and $d=\langle k \rangle/(N-1)$ is the link density of the network. Here the following approximations hold in the limit $N\rightarrow \infty$,
\begin{equation*}
\begin{array}{lcl}
\sum_{j \in S_A}k_j \simeq (1-p)dN^2,~~~~ \sum_{j \in S_I}k_j \simeq pdN^2,\\\\
\frac{1}{N}\sum_{j\in S_A}\frac{d_j^2}{d_j-\alpha_j/\epsilon} \simeq (1-p)F(\epsilon,a),\\\\
\frac{1}{N}\sum_{j\in S_I}\frac{d_j^2}{d_j-\alpha_j/\epsilon} \simeq pF(\epsilon,-b),\\\\
\frac{1}{N}\sum_{j\in S_A}\frac{d_j}{d_j-\alpha_j/\epsilon} \simeq (1-p)\bar{F}(\epsilon,a),\\\\
\frac{1}{N}\sum_{j\in S_I}\frac{d_j}{d_j-\alpha_j/\epsilon} \simeq p\bar{F}(\epsilon,-b),
\end{array}
\end{equation*}
where $$F(\epsilon,\alpha)\simeq \frac{1}{N}\sum_{j=1}^N\frac{d_j^2}{d_j-\alpha/\epsilon},~~~~ \bar{F}(\epsilon,\alpha)\simeq \frac{1}{N}\sum_{j=1}^N\frac{d_j}{d_j-\alpha/\epsilon}.$$
Therefore we obtain
$$\small{J_0=\begin{pmatrix}
\frac{(1-p)}{d}(F(\epsilon,a)+\frac{\eta}{\epsilon}\bar{F}(\epsilon,a)) & \frac{p}{d}(F(\epsilon,a)+\frac{\eta}{\epsilon}\bar{F}(\epsilon,a)) \\\\
\frac{(1-p)}{d}(F(\epsilon,-b)+\frac{\eta}{\epsilon}\bar{F}(\epsilon,-b)) & \frac{p}{d}(F(\epsilon,-b)+\frac{\eta}{\epsilon}\bar{F}(\epsilon,-b))
\end{pmatrix}}$$

The fixed point changes its stability at the phase transition point at $R_A=R_I=0$, which compels us to obtain the following critical inactivation ratio
\begin{equation}
\begin{array}{lcl} \label{eq129}
p_c = \frac{F(\epsilon,a)+\frac{\eta}{\epsilon}\bar{F}(\epsilon,a)-d}{[F(\epsilon,a)-F(\epsilon,-b)]+\frac{\eta}{\epsilon}[\bar{F}(\epsilon,a)-\bar{F}(\epsilon,-b)]},
\end{array}
\end{equation}
 for $\epsilon > \epsilon_c(=a/d_{min})$, where $d_{min} = k_{min}/N$. Particularly, for $\eta = 0$, i.e., when there is no feedback in the system, the critical inactivation ratio becomes  $$p_c = \frac{F(\epsilon,a)-d}{F(\epsilon,a)-F(\epsilon,-b)}.$$ 

  \begin{figure}[ht]
 	\centerline{
 		\includegraphics[scale=0.48]{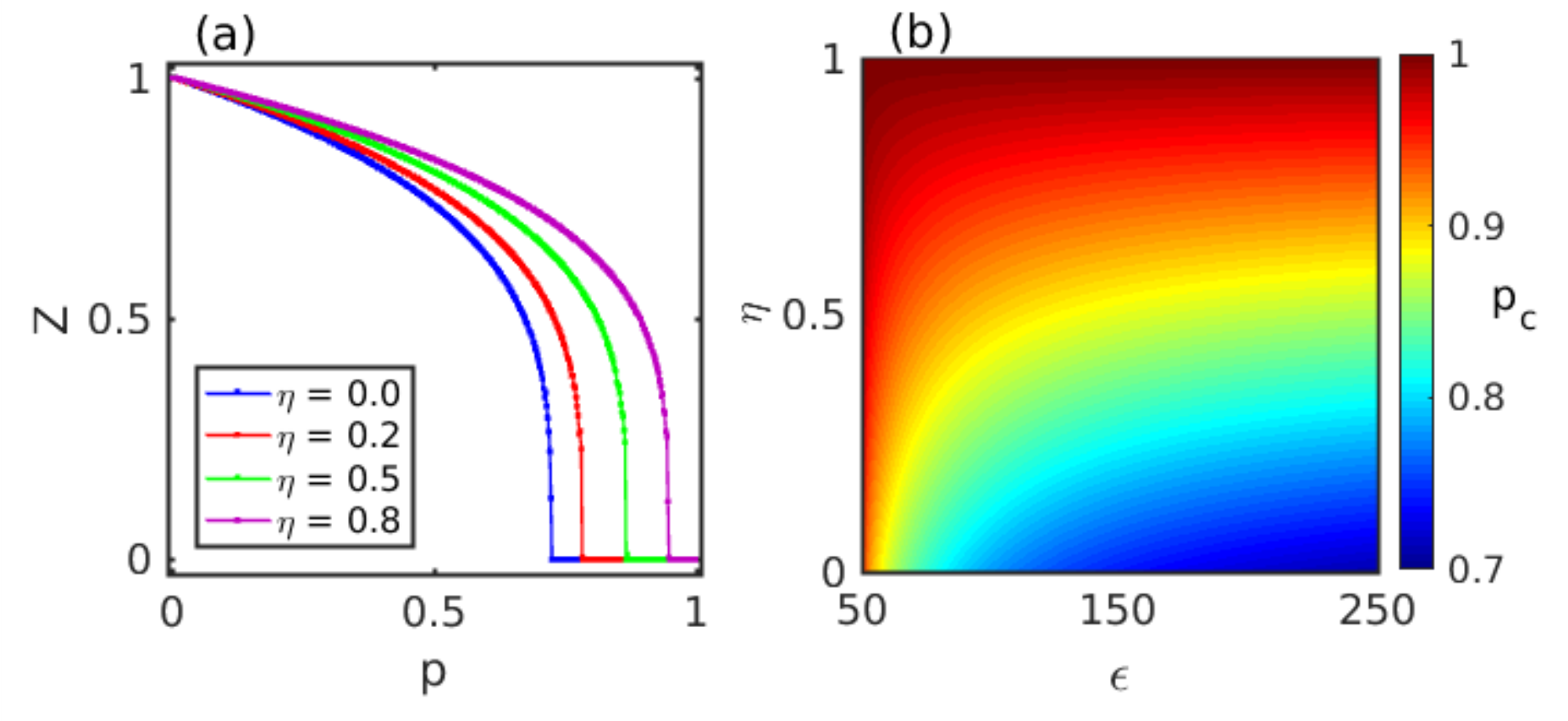}}
 	\caption{Effect of feedback parameter in scale-free networks: (a) The order parameter $Z$ against the inactivation ratio $p$ for various $\eta$ values in scale-free network with $d = 0.078$ and $\epsilon = 200$, and (b) dependence of the critical ratio $p_c$ in $\epsilon-\eta$ parameter space.}
 	\label{fig10}
 \end{figure}
 
 \par  We also plot the critical ratio $p_c$ (cf. Eqn. \ref{eq129}) as a function of $\epsilon \in [50,~250]$ and $\eta \in [0,~1]$ for scale-free configuration of the network in  Fig. \ref{fig10}(b) with the same network and system parameters used for the numerical simulations. Analytically found critical values perfectly match the numerical ones, besides, external feedback has been observed to develop dynamical survivability throughout all values of $\epsilon$.

\section{Effect of feedback on interacting R\"{o}ssler systems}
Finally, we examine the effectiveness of our approach while choosing a chaotic dynamical system coupled through both regular(global) and complex topologies.  
Mathematical form of the damaged network of $N$ interacting delay coupled R\"{o}ssler systems is as follows 
\begin{equation}
\small{	\begin{array}{lcl} \label{eq128}
\dot{x}_j = -y_j - z_j + \frac{\epsilon}{N}\sum\limits_{k=1, k\neq j}^NA_{jk}(x_k(t-\tau)-x_j) + \frac{\eta}{N}\sum\limits_{k=1}^N x_k,\\
\dot{y}_j = x_j + c_jy_j + \frac{\epsilon}{N}\sum\limits_{k=1, k\neq j}^NA_{jk}(y_k(t-\tau)-y_j) + \frac{\eta}{N}\sum\limits_{k=1}^N y_k,\\
\dot{z}_j = d_j + z_j(x_j - e_j) + \frac{\epsilon}{N}\sum\limits_{k=1, k\neq j}^NA_{jk}(z_k(t-\tau)-z_j) + \frac{\eta}{N}\sum\limits_{k=1}^N z_k,

\end{array}}
\end{equation}
for $j= 1,2,...,N$. Here $c_j = d_j = 0.2$, $e_j = 5.7$ for active (chaotic) group of oscillators and $c_j = d_j = -0.2$, $e_j = 2.5$ for inactive oscillators.

\begin{figure}[ht]
	\centerline{
		\includegraphics[scale=0.4]{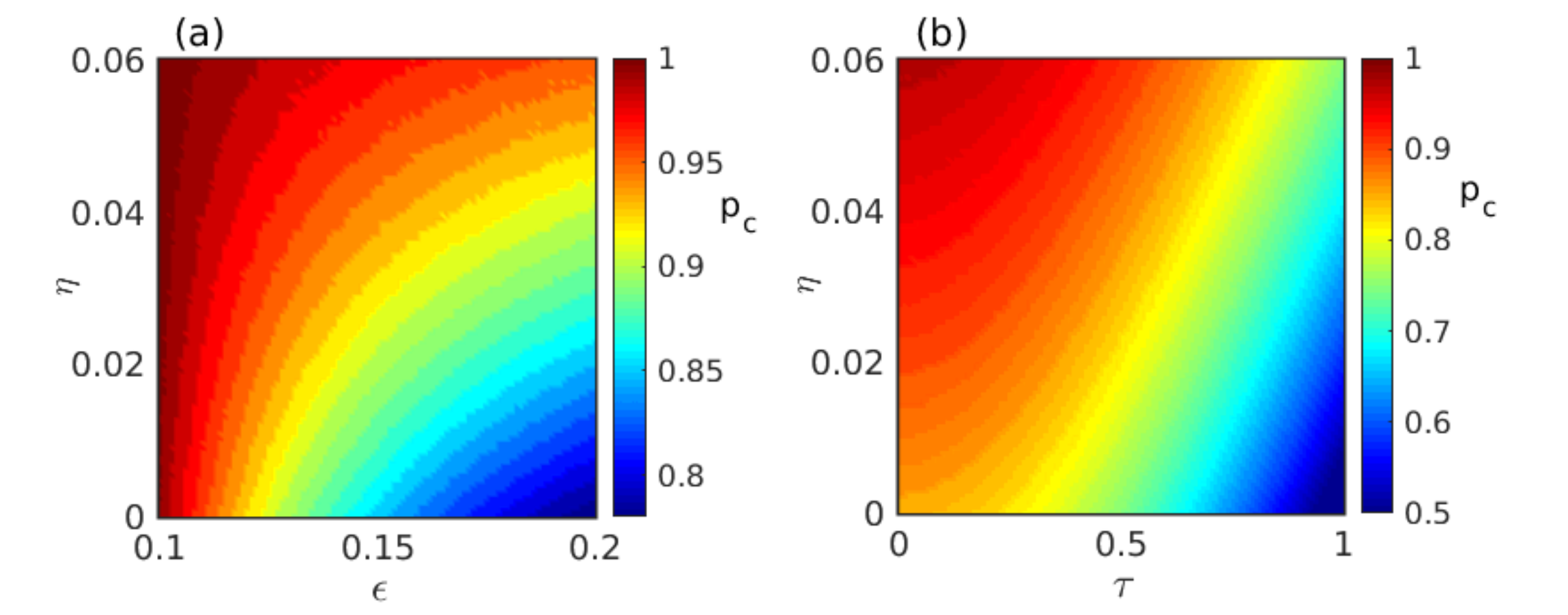}}
	\caption{Dependence of the critical ratio $p_c$ on (a) $\epsilon-\eta$ parameter space for non-delayed interaction($\tau = 0$), and (b) $\tau-\eta$ paramater space for delayed interaction with fixed coupling strength $\epsilon = 0.15$ in a network of $N=500$ globally coupled chaotic R\"{o}ssler oscillators.}
	\label{fig11}
\end{figure}

 We start with defining the order parameter $M$ as $M = \sqrt{\langle({\bf X}_c - \langle {\bf X}_c \rangle)^2\rangle}$, where ${\bf X}_c = \frac{1}{N}\sum_{j=1}^N(x_j,y_j,z_j)$ is the centroid and the bracket $\langle...\rangle$ means a long time average. Aging transition of the system (\ref{eq128}) is further described in terms of this $M$. 
\par Figure 11(a) depicts change in $p_c$ for non-delayed case i.e., $\tau = 0$, with respect to simultaneous variation in the coupling strength $\epsilon \in [0.1,~0.2]$ and feedback strength $\eta \in [0,~0.06]$. Whenever $\eta=0$, $p_c$ gradually decreases for increasing $\epsilon$ and reaches $p_c\simeq 0.78$ for $\epsilon=0.2$. But, as we employ a feeble increment in $\eta$, it starts raising this critical value of $p$ greatly even to $p_c\simeq 0.95$ for $\eta=0.06$. This scenario is valid for any value of $\epsilon$ implying significant improvement in the network resilience. The impact of $\eta$ in case of delay coupled oscillators is also illustrated in Fig. 11(b) for time delay $\tau \in [0,~1]$. Moreover, to corroborate the generality of our approach, in Fig. 12 we unveil the effect of feedback parameter $\eta$ on a network of coupled chaotic R\"{o}ssler oscillators with small-world and scale-free topologies. Figure 12(a) portrays the dependence of the critical ratio $p_c$ on simultaneous variation of the coupling strength $\epsilon \in [0.4,~1]$ and the feedback strength $\eta \in [0,~0.06]$ in a small-world network with rewiring probability $p_{sw} = 0.01$ and link density $d = 0.24$. Whereas, Fig. 12(b) depicts the result for variations of $\epsilon \in [2.5,~5]$ and $\eta \in [0,~0.06]$ in a scale-free network with link density $d = 0.078$. Noticeably, for both these architectures, the applied feedback enhances the network survivability against badness of the dynamical units. This is how the proposed mechanism of enhancing survivability for time-delayed and complex networks works whenever chaotic systems are used to cast the active units in the damaged network.

\begin{figure}[ht]
	\centerline{
		\includegraphics[scale=0.4]{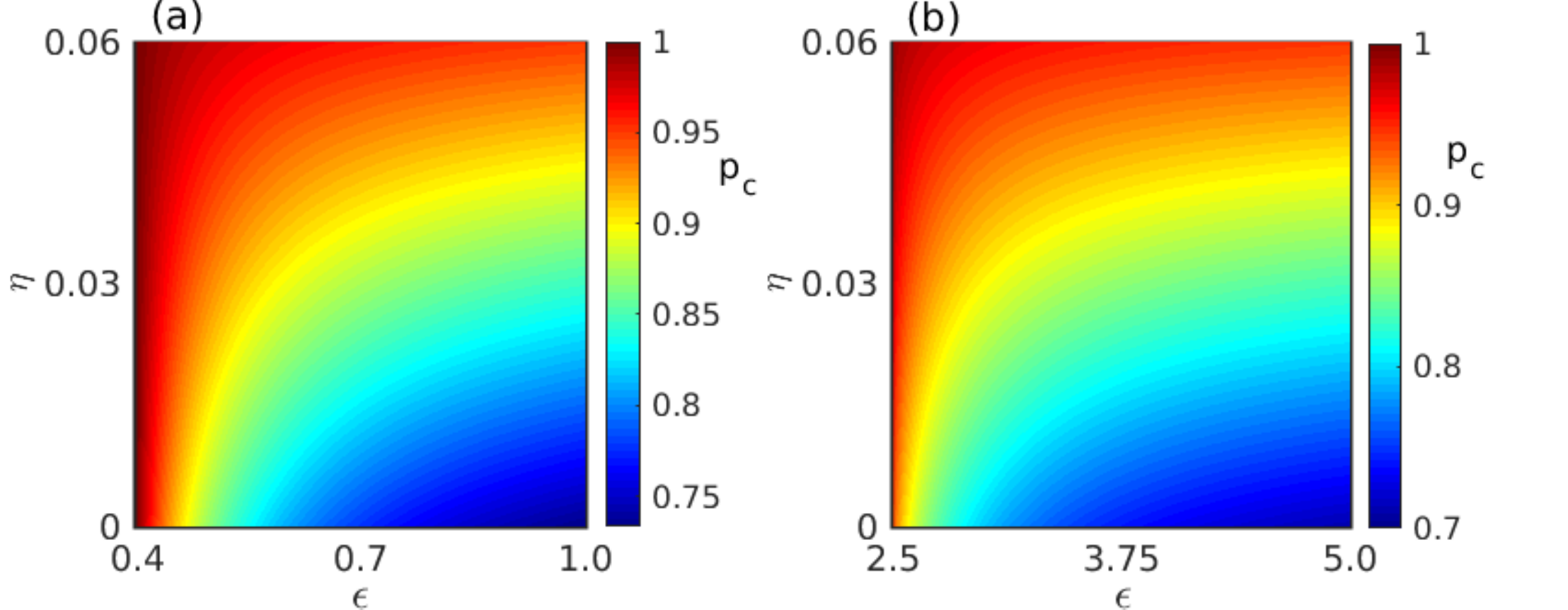}}
	\caption{Dependence of the critical ratio $p_c$ on coupling strength $\epsilon$ and feedback parameter $\eta$ considering (a) small-world, and (b) scale-free network of $N=500$ coupled chaotic R\"{o}ssler oscillators.}
	\label{fig12}
\end{figure}

\section{Conclusions}
Network robustness has recently become a topic of great research interest because it bears resemblance to various natural occurrences. In the present article, we have gone through the notion of rescuing networks from a throughout collapse due to a specific type of dynamical perturbation. Particularly, we have examined the dynamical robustness of damaged networks in terms of aging transition and demonstrated how one can resume dynamism through feedback mechanism. We have shown that adding simply an appropriate  linear feedback term in the aging network, network's survivability can be developed quite substantially. This enhancing impact of the feedback is observed to be effective irrespective of the coupling strength, no matter how large it is. We have illustrated this scenario through both numerical and analytical findings while considering limit cycle Stuart-Landau systems as the local dynamical units. A comparative study on the process of adding feedbacks to only active, solely inactive and both active-inactive groups of nodes has also been presented. Quite remarkably, we have been able to develop network persistence  even in the presence of time delay in the interaction among the nodes, although delay is a candidate that may effectively lower the network resilience. As far as the generalization of our approach over different network topologies is concerned, we have realized same qualitative features of external feedback in small-world as well as scale-free complex structures of the aging networks. In order to reveal the local system independence of our scheme, we successfully performed similar analysis for chaotic R\"{o}ssler oscillators. Our study may have exclusive applications in increasing the survivability of several natural systems that experiences local inactivation of its components.

\medskip

\par {\bf Acknowledgments:}
D.G. was supported by SERB-DST (Department of Science and Technology), Government of India (Project no. EMR/2016/001039).


\begin{thebibliography}{25}
\bibitem{mthd} The fifth order Runge-Kutta Fehlberg scheme and modified Heun's method have been used to integrate the non-delayed and delayed dynamical equations respectively.
\bibitem{op} The numerical results are calculated using the same order parameter as defined in section II.
\end{thebibliography}
\end{document}